\documentclass[aps,showpacs,preprintnumbers,amsmath,amssymb, 
twocolumn, tightenlines,
]{revtex4}

\usepackage[dvips]{graphicx}\usepackage{bm}
\usepackage{amsmath}
\usepackage{dcolumn}
\usepackage{bm}

\newcommand{\be}{\begin{eqnarray}}
\newcommand{\ee}{\end{eqnarray}}

\begin{document}

\title{Influence of Meson's Widths on\\
 Yukawa-like Potentials and Lattice Correlation Functions}
\author{ V.V. Flambaum$^1$ and E.V. Shuryak$^2$ }

\affiliation{$^1$
 School of Physics, The University of New South Wales, Sydney NSW 2052,
Australia}

\affiliation{$^2$ 
Department of Physics and Astronomy, State University of New York, 
Stony Brook NY 11794-3800, USA
}

\date{\today}

\begin{abstract}
Euclidean point-to-point propagators or wall-to-wall correlators
related to exchange by an unstable particle
($\sigma,\rho,\omega$-mesons)
are modified by presence of particle width. In particular, the
usual method of deriving particle masses from logarithmic derivatives
need to be modified.  
Similarly  Yukawa-like potentials of nuclear physics due to exchange 
of those mesons  are significantly modified
since the coupling to the decay products is strong.
For example, the large distance asymptotic changes, $\sim \exp(-M_{min}r)$,
where $M_{min}$ is the sum of the decay product masses
 ($2 m_{\pi},2 m_{e}, 2 m_{\nu}$). In the area $M_{min}r<1$ the potential
 has a long-range tail $\sim 1/r^3$. Similar effects appear due to the virtual
decays in the elecroweak sector of the Standard model. 
The $Z-\gamma$ mixing via electron loop  gives the parity violation
potential with the range $1/2 m_{e}$, i.e. the range of the weak
interaction increases $10^{5}$ times.
\end{abstract}

\vspace{0.1in}

\maketitle

\section{Introduction}
  This paper addresses the following  general question:
  what is the role, if any, of  a  width $\Gamma$ of an 
unstable particle
in  situations in which this particle is a $virtual$ intermediate state
so that its physical decay $cannot$ take place.
We will discuss three applications of such kind: \\
 (i) Euclidean $point-to-point$ correlation functions in which
  particles in question are 
 emitted and absorbed by  two  operators acting at 4-d points;\\
  (ii) Euclidean $wall-to-wall$ correlators, widely  used on the lattice for determination of particle masses; in this case a signal is emitted and absorbed
  from 3-d planes separated by some time interval
$\tau$;\\
(iii) Yukawa-like potentials due to meson exchanges widely used   in nuclear physics,
this case can be called $line-to-line$ correlators.

\begin{figure}
 \includegraphics[width=7cm]{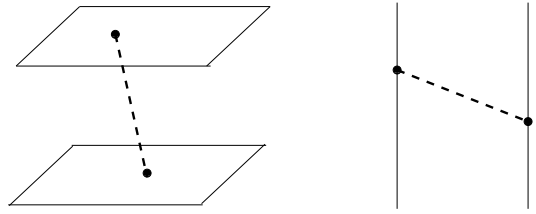}
 \caption{\label{fig_corr}
 The setting for (a) wall-to-wall and (b) line-to-like
Euclidean correlators, describing Euclidean lattice correlators and
exchange potentials, respectively. The unstable article propagation is
indicated by a dashed line.
 }
 \end{figure}

The cases (ii) and (iii) are schematically illustrated in Fig.\ref{fig_corr}:
although those have direct applications, the corresponding
results simply follow from those  of the generic case (i), the
point-to-point Euclidean propagator. 
There are of course other similar situations in which the same
ideas apply but not discussed below: t-channel meson exchange in  scattering
amplitudes is another example. There would be no difficulty to generalize what is
 done below to these case as well. Also let us add that
although we call unstable particle  
   a meson, it can equally well be any other unstable particle with
   non-negligible width
  (e.g. the $Z$ boson of weak interaction).
   
 To set up a problem, let us first explain why naive Euclidean continuation of
the real-time amplitudes does $not$ work. Usual real-time amplitudes
describing resonances behave as damped oscillators
\be A(t)\sim exp(-iMt-\Gamma t/2) \ee 
Their naive Euclidean rotation, by simple substitution $t\rightarrow
-i\tau$
which leads to an oscillating function $\sim exp(-M\tau +i\Gamma
\tau/2)$ which certainly
cannot be true Euclidean propagators or amplitudes. 
In fact those are defined via real exponent of the Hamiltonian
$\sim exp(-\hat H \tau)$ and standard
spectral decomposition plus Hermitian Hamiltonian
 of course demands all energy to be real and thus the correlator 
 to  decrease monotonously 
with $\tau$.

 Why should one seek an answer to such questions?
Apart of being the basis of applications of statistical mechanics,
  Euclidean correlation functions with real exponentials
of $\hat H$ are the main tool of  lattice
gauge theory community,  which
studies the non-perturbative dynamics of non-Abelian gauge fields
by numerical simulations, see multiple reviews e.g. \cite{lat_revew}
. Rotation to Euclidean time is also a crucial
ingredient of the
 semiclassical approach  based on instantons, see e.g.review\cite{inst_rev}.
These correlators are widely used to extract the particle masses and their
couplings to certain currents, but  it was not yet
used to extract the {\em widths} of resonances\footnote{Only one very
  recent
paper \cite{Cristoforetti:2007ak} has correlators
 (calculated using the instanton liquid model) with such high
statistical quality
that
 discussion of the magnitude of the rho meson width became
meaningful.}. We will see below that
it is indeed not a simple task.

 The present-day lattice calculations have practical
limitations stemming
from available computer resources. As a result, quark used
are still rather heavy, so that in most calculations  the 
pions are so heavy that   no decays of $\sigma$ and $\rho$ are possible.
Another kinematical effects is a consequence 
of  small  size of the present-day lattices $L\sim 2$ fm. 
Indeed,  momentum quantization of the decay products due
to the periodic
boundary
conditions  to $p_n=2\pi n/ L$ also prevents decay in
many cases, especially with the  
the initial ``plane-to-plane'' projection\footnote{
Even wide sigma mesons is not decaying,
see e.g.\cite{Mathur:2006bs}.
A way to partially relieve this problem is not to project to $\vec p=0$
and use well-localized sources, with many momenta included.
} to $\vec p=0$.

  This paper is not addressing these lattice limitations,
considering ideal case with quark masses at its physical value
and  $L$  large enough to cause no problem.
The issue we address
is how a propagator of unstable particle look like, when
it is virtual and is propagating either in space direction
(related to Yukawa
potentials below) or in imaginary (Euclidean) time.
As we will see, 
the widths  can modify the shape of these propagators
, sometimes beyond recognition. Although masses and widths
can still be extracted from those correlators, it would  require
 very complicated fitting procedure. For the same reason, using lattice
simulations for the evaluation
 of the transport properties  via Kubo-like
formulae from Euclidean correlators is practically impossible, 
as discussed e.g. in
\cite{PetTeaney}.

  Our approach  is based on  standard dispersion relations, which
connect
  real part of the amplitudes
in momentum representation to their imaginary part, also known as
 {\em spectral densities},
  in which resonances appear as Breit-Wigner-like peaks. 
  If the resonance width is very small, those can be approximated by 
delta functions and the width ignored.
If so, the usual exponential behavior of the wall-to-wall Euclidean
 correlators $K(\tau)\sim exp(-M\tau)$
or standard Yukawa-like potentials $V(r)\sim exp(-Mr)/r$ are obtained.
 If the width is small,
one can perform a saddle point integration and find analytically 
the first corrections to delta function, in form
of some ``shifted mass" depending on the width $\Gamma$.
 If the width is not small, 
the form of the correlators or potentials is strongly
modified, as it will be detailed below.  

Another related issue  is 
 modification of hadronic resonances
at finite temperatures. We will not discuss here
multiple physical effects which are important in resonance
emission from ``heat baths'' created e.g. in heavy ion collisions,
but just comment on the simplest ``kinematical'' effect 
\cite{BrownShu} which appear even in high energy 
elementary    $pp$ collisions.
When the spectral density is convoluted with 
the thermal Boltzmann exponent,  the finite width of a resonance
  leads to different weighting of the left and right
side of the peak and thus an apparent resonance mass
shift. For small width  and/or high temperature it is very easy to
estimate the effect: approximating the resonance as
\be {exp(-M/T)\over 1+(4/\Gamma^2)(M-m_{res})^2}\approx \ee
$$ exp[-({4\over \Gamma^2})(M^2-2M m_{res}-M^2+{2M\Gamma^2\over 8T}]
$$ 
one gets the
mass shift to be 
\be \label{eqn_T_shift}
\delta m_{max}\approx m_0 -{\Gamma^2 \over 8 T}\ee
This simple analytic expression however does not work well in practice
and real shift depends much on the resonance
shape (see \cite{BrownShu} for details).

 \section{Point-to-point correlators, dispersion relations and the
   spectral density}
 We start with the point-to-point  correlation functions 
because other cases can easily be obtained from them, by
integration. In fact point-to-point correlators
in Euclidean
 space-time  are not just methodical object, they are important  
  tools widely used in studies of 
 structure of the QCD vacuum.  They can be 
deduced phenomenologically, using 
vast set of data accumulated in hadronic physics.
Second, they can be directly 
calculated {\em ab initio}  using quantum field theory methods, such as lattice gauge
 theory, or semiclassical methods. 
 Significant amount of work has also been
done in order to understand their small-distance behavior,  based
on the Operator Product Expansion, see \cite{SVZ_79} and vast
subsequent literature.

These correlation functions are   vacuum expectation value  of
the product
of two (or more) of operators 
\be K(x-y)= <0| O(x) O(y)|0 > \ee
which can have any gauge invariant combination of fields, for example
mesonic operators we will consider have local quark-antiquark 
combination 
\be O_{mes}(x)=\bar\psi_i M_{ij} \psi_j \ee 
where a matrix M can include various flavor and  Dirac structures,
and a convolution over quark color indices.
Since
 the vacuum is homogeneous,  the correlators depend on the 
relative distance only. If two points are inside the light cone,
a lot of things may happen: e.g. a meson may decay in between.
We will not consider such case below, always assuming that
the distance between the points  (x-y) is  {\it space-like}.
It may really be a space distance or ``Euclidean time'' interval:
in both cases there are no real decays and  
one deals with monotonously  decreasing  functions of  the distance.

 The Fourier transform of $K(x)$, 
$K_{mom}(q^2)$,
 depends
on the momentum transfer $q$ flowing from one operator to another. 
In its language, the situation we consider corresponds only to
 {\it virtual}
 space-like 4-momenta $q$
 $Q^2=-q^2>0$, 
 like in scattering experiments in which the resonance is
 in $t$ and not $s$ channels.

 Due to causality, 
  standard dispersion relations follow: they 
 relate real and imaginary parts
\be \label{eq_disp_rel}
K_{mom}(q^2)= \int {ds\over \pi} {ImK_{mom}(s) \over (s-q^2)} \ee
where the r.h.s.  is the {\it physical spectral density}
$ImK_{mom}(s)$. It  certainly
is non-zero only for  {\it positive} $s$ above certain threshold
(twice the pion mass in the case of $\rho,\sigma$ resonances to be 
considered\footnote{Because
we  only consider {\it negative} $q^2$, in the semi-plane 
without singularities, we never come across a
 vanishing denominator and therefore  Feynman's  $i\epsilon$
  in the denominator of the pole is not indicated. }.

   Dispersion relation is the basis of various {\it
sum rules}. Their general idea is
as follows; suppose one knows 
the l.h.s. $K_{mom}(q^2)$ somewhere; thus
  some integral in the r.h.s. of 
the physical spectral density is known as well. 
(For example, for mesonic operators the correlators at small x
  are just given by free quark propagation $K\sim 1/x^6$, with
  calculable corrections.)
The simplest of them 
use directly momentum space: those are known as
 {\it finite energy sum rules}. Unfortunately, those 
are not very useful because  most of the dispersion 
integrals are divergent and  to usable 
  sum rules appear only after some ``subtractions'', which
 introduces extra parameters and undermine their prediction 
power.

  Improved  sum rules appear if one
takes a
sufficient number of derivatives of the dispersion relation
at $Q=0$,  defining the so called
{\it moments} of the spectral density
\be M_n=  \int {ds\over \pi} {Im K_{mom}(s)\over  s^{n+1}} \ee
Following Shifman,Vainshtein and Zakharov \cite{SVZ_79},
this method is traditionally used in
the discussion of ``charmonium sum rules" related to correlators 
of $\bar c c$ currents.
  Another idea also suggested by these authors 
 \cite{SVZ_79}
is to introduce
 the so called {\it Borel transform } of the function $K_{mom}(Q)$
 defined as follows
\be K_{bor}(m)= lim_{n,Q^2 \rightarrow \infty,m^2=Q^2/n^2=fixed}
\nonumber \\ 
{Q^{2n} \over (n-1)!} (-{d \over dQ^2})^n K_{mom}(Q^2) \ee 
leading to ``Borel-transformed'' sum rules
\be K_{bor}(m)= \int  {ds\over \pi} Im K_{mom}(s) exp(-{ s\over m^2}) \ee 
in which the integral is cut off at large $s$  exponentially. 
Such form of the sum rules have been widely used 
in multiple papers based on the QCD sum rules.

   Another natural option, advocated by one of us in \cite{Shu_cor},
is to use the sum rules directly in the
(Euclidean)  coordinate representation.
 By applying Fourier transformation to (\ref{eq_disp_rel})
one obtains the following 
nearly self-explanatory form 
\be \label{eqn_corr}
K(x) =  \int  {ds\over \pi} Im K_{mom}(s) D(s^{1/2},x) \ee 
The  former function in the r.h.s. describes 
the amplitude of production
of all intermediate states of mass $\sqrt{s}$, while the latter 
\be D(m,x)=(m/4 \pi^2 x) K_1(mx) \ee 
is nothing else  but the Euclidean propagator of a mass-$s$  state 
 to the distance 
$x$.  At large $x$ the propagator goes as $exp(-mx)$, thus
 it is also exponentially cut off and, in practice, there
 is little difference between this expression and Borel sum rules
. However
the coordinate form can be much easier evaluated
numerically, on the lattice or in semiclassical models,
unlike the Borel-transformed version. As discussed in \cite{Shu_cor}
and subsequent literature, for vector and axial correlators
one can directly use experimental data and calculate
the l.h.s. with rather small error bars. This can provide
a very good test of lattice calculations. 
The  inverse transformation -- from correlators to spectral densities
 --
is a badly defined problem, quite difficult in practice, except in
some simplest situations. 

Now, let us think about a situation in which the lowest state
in the corresponding channel is an $unstable$ particle -- for example
$\rho,\omega$ mesons for vector  operators or $\sigma$ for
scalar isoscalar operator $\bar q q$. In the simplest case of very
narrow resonance
in which the width of the resonance is neglected, one simply gets
the propagator of a state with the resonance mass $M$
 $ K(x) \sim D(M,x) $. If the width is small, one can use
the same idea as for the thermal case, leading to
analog of the formula (\ref{eqn_T_shift}) with the temperature
$T$ replaced by $1/x$. If the width is substantial,
the integral in (\ref{eqn_corr}) should be calculated more accurately,
with
the appropriate shape of the resonance. We will show how this
procedure
works in the next section.

\section{Wall-to-wall correlators and lattice measurements of hadronic masses}
 Although there is quite substantial literature on lattice
 point-to-point correlators, most of studies rely on
{\it wall-to-wall} correlation functions,  obtained from K(x) by 
additional integration
over the 3-dimensional spacial
plane. This
puts the   momentum of all intermediate   states
to be zero, so  dispersion relation in this case can be
considered to be related not to the mass of the state but 
simply to its energy.  The resulting
function is
related to the physical spectral density by
\be  \label{Ktau} K_{wall-to-wall}(\tau) =  \int {dm\over \pi} Im K_{mom}(m) exp(-\tau m) \ee 
where 4-d propagator $D$ is substituted by its 1-d version.

 The mass of the lowest hadrons is traditionally obtained 
from the logarithmic derivative of such correlators 
\be \label{Etau} E(\tau)= -\frac{d log K}{d \tau}\rightarrow_{\tau\rightarrow\infty} M    \ee 
Indeed, multiple lattice calculations have been able to
show that
the logarithmic derivative of wall-to-wall correlators
do have  ``plateaus'' independent of $\tau$,
being the standard source of the meson masses.

  Such procedure works well in present-day calculations in which
kinematical effects mentioned above prevent decays.
However, as the algorithms and
supercomputers gets more powerful,  
 the widths eventually would become an issue.

\begin{figure}
 \includegraphics[width=7cm]{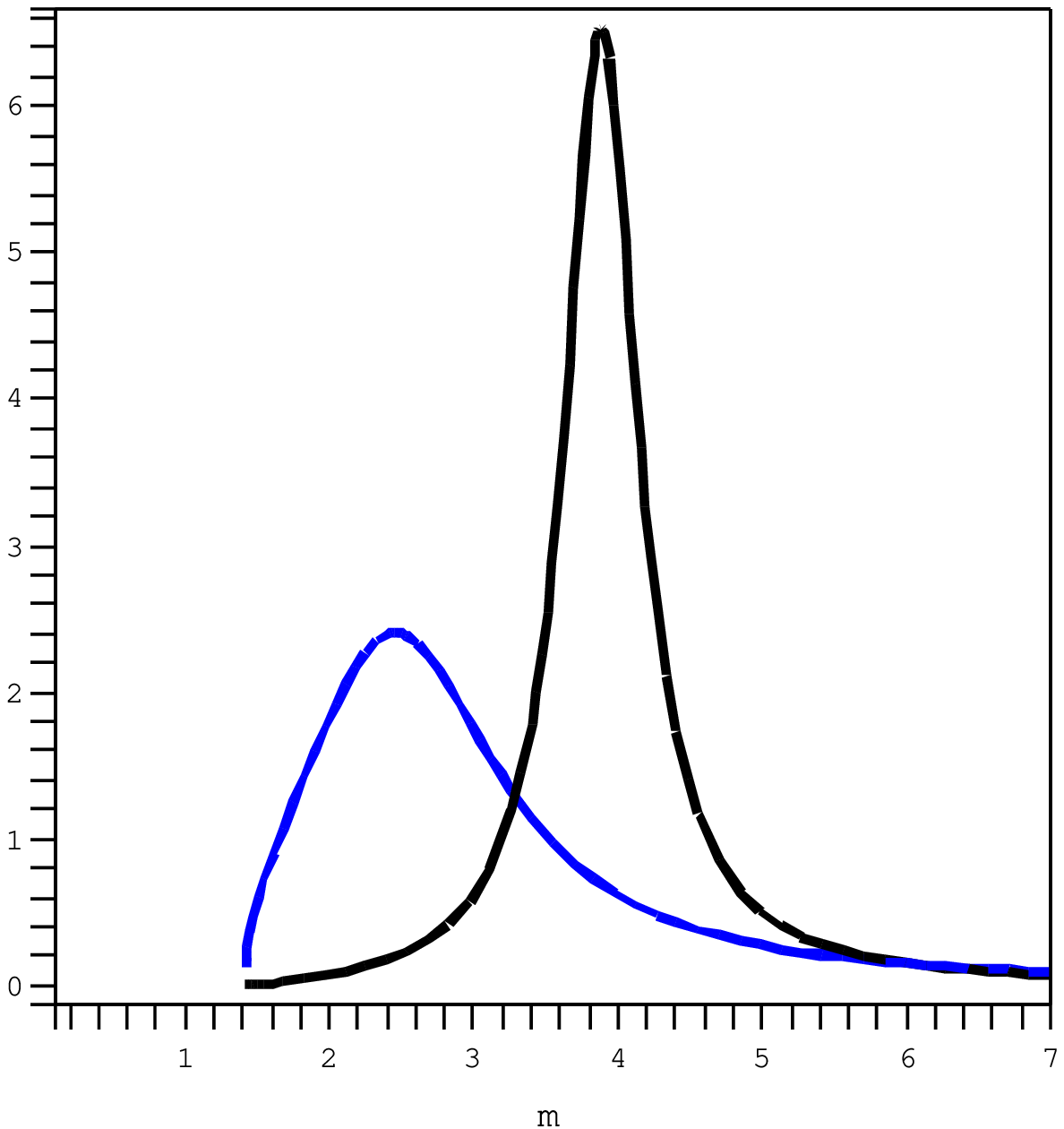}
 \includegraphics[width=7cm]{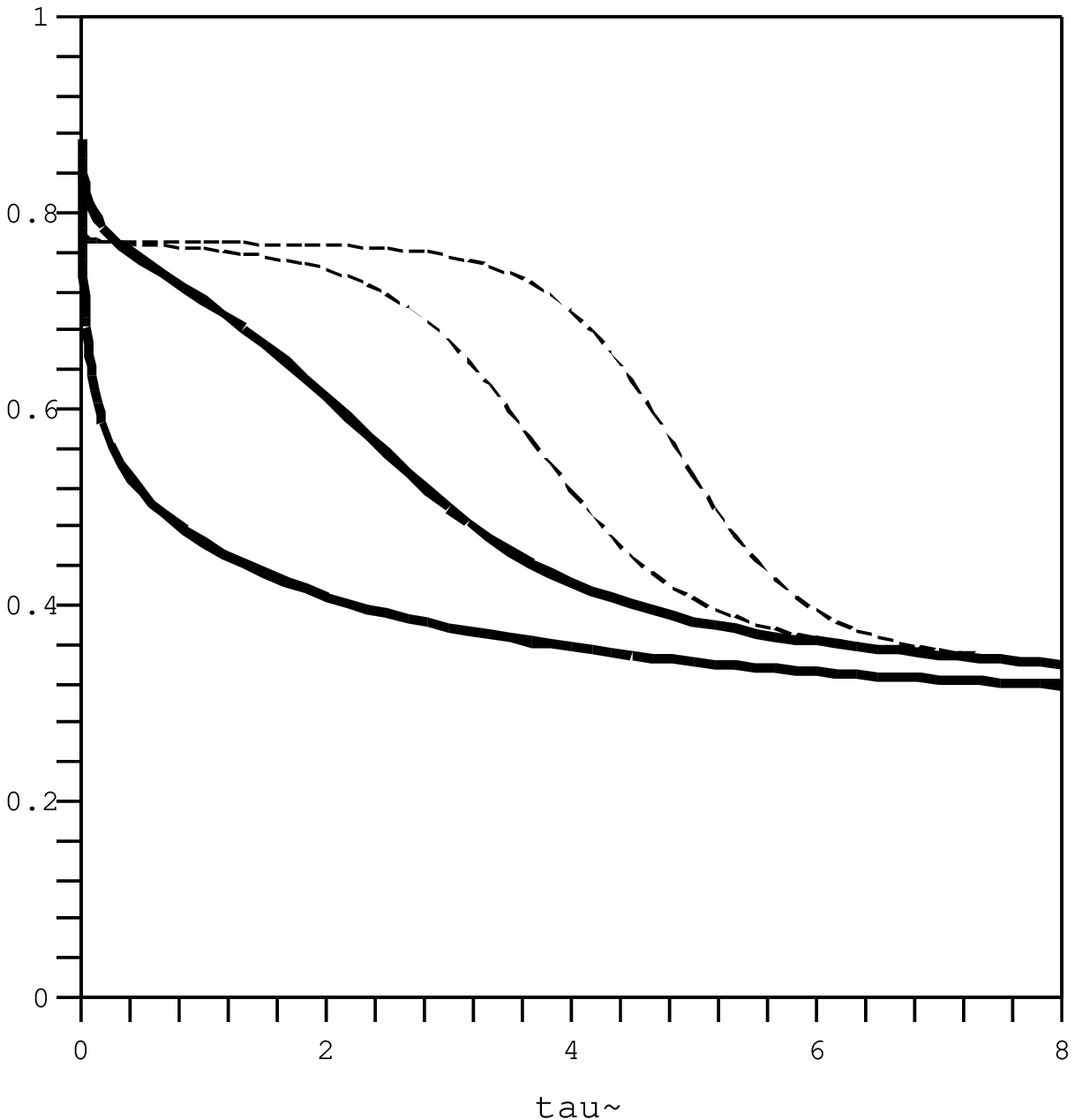}
 \caption{\label{fig_rho}
(a) The shape of $\rho$ (right) and $\sigma$ resonance, as a function
   of its mass m ($fm^{-1}$). (b) 
The corresponding logarithmic derivative of the 
wall-to-wall correlator $E(\tau)\, (GeV)$
vs Euclidean time $\tau (fm)$. The solid curves are for $\rho$
(higher)
and $\sigma$ resonances (both mass and width
are 450, MeV), two dashed  curves are for $\rho$ with the width
reduced to $\Gamma_0=15,1.5 \, MeV$, (lower and upper, respectively).
 }
 \end{figure}

  Let us demonstrate what should happen with ``mass plateaus'' 
by calculating them for different resonance shape. The simplest
case in which an analytic formula is easily obtained is a Gaussian
resonance $\sim exp[-4*(m-m0)^2/\Gamma^2]$ for which
the mass integral can be taken from minus to plus infinity and 
\be E(\tau)=m-\tau \Gamma^2/8 \ee
But this dependence does not reproduce what happens for
realistic shapes of the resonances, with a particular
dependence of the width on energy, especially near the
threshold. The best way is to calculate 
numerically the behavior of $E(\tau)$
for realistic spectral densities, and see what should its true shape 
 for vector and scalar currents be. 
 The most studied  vector currents have two
lowest resonances, I=1 $\rho$ meson and I=0
$\omega$. In principle, the spectral densities
describe continuum of states, all the way from
 the lowest physical states
 (2 and 3 pions at rest) respectively. So, one may only argue that
the logarithmic derivatives is limited by
  $2m_{\pi}$ or $3m_{\pi}$ from below. 

A reasonably good 
parameterization of the spectral density $Im K_{mom}$ for binary channels is
Breit-Wigner
\be \rho(M)=   { \Gamma(M) \over (M-m_{res})^2+\Gamma(M)^2/4}
\ee
where the width is not constant and depends on the
current running mass $M$ and  angular momentum of the resulting pions.
Due to P-wave nature of $\rho\rightarrow \pi\pi$
decay, its $\Gamma_\rho(M)=\Gamma_\rho^0 (1-4m_\pi^2/M^2)^{3/2}$,
while for $s$-wave $\sigma$
resonance one should use the power $1/2$ instead.
The corresponding shape of the resonances is shown in
Fig.\ref{fig_rho}(a): we do not show here fitting of the real
data, although we have checked that these shapes do indeed 
 agree with them.

The resulting 
``realistic'' logarithmic derivative of the meson contribution
is shown in Fig.\ref{fig_rho}(b). Let us start
its discussion with two dashed  curves, which  are for
$\rho$ meson with the width artificially reduced by the factor
100 (upper dashed) and 10 (lower dashed curve)
compared to real-world $\Gamma_0=.14 \, GeV$. These
two curves
nicely show ``plateaus'' at the rho mass $M_\rho=.77\, GeV$
at not-too-large $\tau$, with
a  ``slide'' 
toward   the threshold ($2m_\pi$)
 at large $\tau$. However for the realistic width
value (the solid curve) the rho meson mass plateau
is already gone, with a slide  of $E(\tau)$
 even at small $\tau$.
The situation is even more dramatic for the sigma meson,
for which the mass-to-width ratio is 1:1.
Presumably by fitting such curves, if they come
from correlator studues, one would still be able to recover both 
the mass and the  meson width, but we leave this for further studies.

     It may be useful to have some analytical estimates for $K(\tau)$.
Let us start from  large $\tau$. In this case small $m$ dominate
in the integral (\ref{Ktau}) and we can substitute $m=2 m_{\pi} (1+x)$
assuming $x<<1$. The result is
\be  \label{Klarge} K(\tau) = const \frac{2 m_{\pi}\Gamma_0}{(m_{res}-2 m_{\pi})^2}
\frac{ exp(-2 m_{\pi}\tau )}{( m_{\pi}\tau )^{q+1}},\\
 E(\tau)= -\frac{d log K}{d \tau}= 2 m_{\pi}+(q+1)/\tau ,\ee 
where $q=3/2$ for $\rho$ and  $q=1/2$ for $\sigma$.
This should be compared to the resonance contribution
$K^r(\tau)= exp(-m_r \tau )$. Only if this term dominates,
we have $E^r (\tau) \approx m_r$. However, for $\rho$ and
$\sigma$ the coefficient $2 m_{\pi}\Gamma_0/(m_{res}-2 m_{\pi})^2$
is not small, therefore, the asymptotic regime starts very early.
 The transitional area corresponds to $ 2 m_{\pi} < 1/\tau <m_r$.
Here an essential  contribution comes from
 the integration interval $ 2 m_{\pi} < m  < m_r$:
$ K^{tr}(\tau)\sim \Gamma_0/(\pi (m_{res}-2 m_{\pi})^2 \tau)$.
If this contribution dominates we have $E^{tr}(\tau) \approx 1/\tau$.
Note that the value of the width  actually determines the position
of the transition point between  the areas of the resonance and non-resonance
domination in  $K(\tau)$ or  $E(\tau)$. Therefore, both the meson mass
and width can, in principle, be found from the fit of lattice calculations.

 \section{Yukawa-like potential due to exchange by an unstable particle}

 Let us consider the potential between two
nucleons produced by a meson (e.g. $\sigma,\rho,\omega$) exchange.
For brevity we omit all isotopic and kinematic structures which are different
for different particles and may be easily restored in the final answers. 
This potential in the coordinate representation may be presented as an integral
of its imaginary part in the momentum representation 
 \begin{equation} \label{Ur}
U(r)=\frac{1}{4 \pi^2 r} \int^{\infty}_{2m}
 {\rm Im} U(M)
 \exp(-r M) 2MdM \ ,
\end{equation} 
One can either consider this expression as simply a line-to-line
variant of the Euclidean correlators, with 3-d Green function
$exp(-Mr)/r$
substituting 4-d one in (\ref{eqn_corr}), or follow the usual
 derivation of the Uehling potential in textbooks 
(such as \cite{berestetskii}).
 The lower limit of the integration is determined by the minimal mass
$2m$ of the decay products ($\pi$-mesons) of the intermediate particle and 
\be \label{UM}
{\rm Im}U(M)=-{\rm Im}\frac{4 \pi g_1 g_2}{M^2-M_0^2-P(M)}=\\
\frac{4 \pi g_1 g_2 m_r \Gamma(M)}{(M^2-M_m^2)^2+
m_r^2 \Gamma^2(M)}\ .
\ee 
Here $g_1, g_2$ are the meson-nucleon interaction constants, $P(M)$
 is the polarization operator,$M_0$ is the bare meson mass,
 $M_m=M_m(M)=\sqrt{M_0^2+{\rm Re}P(M)}$ is the running meson mass,
$m_r=M_m(m_r)=\sqrt{M_0^2+{\rm Re}P(m_r)}$ is the meson mass at
 the resonance, and 
${\rm Im}P(M)\equiv -m_r \Gamma(M)$ where $\Gamma(M)$
is the running meson width.

Substituting Eq. (\ref{UM}) into  Eq.(\ref{Ur}) we obtain
\be\label{UBW}
U(r)=\frac{g_1 g_2 m_r}{ \pi r} \int^{\infty}_{2m}
\frac{\Gamma(M) \exp(-r M)}{(M^2-M_m^2)^2+m_r^2 \Gamma^2(M)}
 dM^2.
\ee 

The explanation of this formula, like other Euclidean
correlators above, is  simple: the mass $M$ of the unstable
particle is not fixed and 
therefore the  potential can be seen as an the integral over all possible
  meson masses given by the spectral density
(e.g. a modified Breit-Wigner formula we use).

\begin{figure}
 \includegraphics[width=7cm]{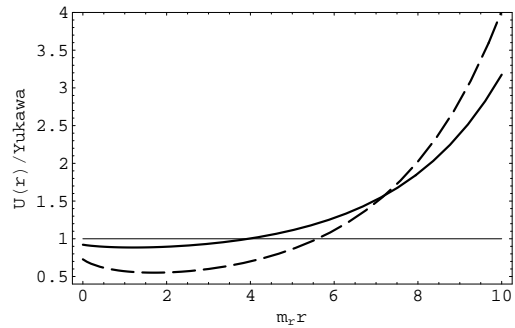}
 \caption{\label{sigma}
The ratios of $\rho$-meson  potential to Yukawa potential
(thick solid line) and  $\sigma$-meson  potential to Yukawa potential
(dashed line) as functions of $m_r r$. At  $m_r r<4$ the negative
corrections given in Eq.(\ref{Ures1}) dominate. At $m_r r>4$
the positive corrections Eqs.(\ref{U2m}-\ref{Up}) change the long-range
behavior of the potentials.}
 \end{figure}

In the case of vanishingly small $\Gamma(M)$ the Breit-Wigner factor
in  Eq. (\ref{UBW}) is proportional to $\delta$-function and the integration
gives the usual Yukawa potential:
\be\label{Yukawa}
U(r)=\frac{g_1 g_2}{r} \exp(-r m_r)
\ee
Non-zero width $\Gamma(M)$ produces significant deviations
from the simple Yukawa formula.  The potential $U(r)$ in Eq.
(\ref{UBW}) can be easily found
using numerical integration. At  Fig.\ref{sigma} we presented graphs of
two ratios: $\rho$-meson potential divided by the Yukawa potential and
 $\sigma$-meson potential divided by the Yukawa potential .
We see that the deviations from the  Yukawa formula may be very large,
 especially at large distances. Below we perform analytical estimates
to explain the nature of different corrections to Yukawa
formula. For the analytical estimates we assume 
$m<<m_r$ and  $\Gamma_r << m_r$ where $\Gamma_r =\Gamma(m_r)$
 is the resonance width.

 The  integral has four important intervals of the integration:
the first interval is near the threshold, $M \sim 2m$; the second interval
is $2m<<M<<m_r$; the third interval is near the resonance, 
 $M \sim m_r$; and the fourth interval is
 $\Gamma_r <<M-m_r<infinity $.
Therefore, the potential may be presented as a sum of four terms
\be\label{Utot}
U(r)=U_{2m}(r)+ U_{p}(r)+U_{res}(r) +U_{inf}(r)
\ee
 Consider first a behavior  of the potential
 (\ref{UBW}) at large distance, $r m_r >> 1$. In this case
the threshold interval $M \sim 2m$ gives the dominating contribution.
The width $\Gamma(M)$ vanishes near the threshold as 
 $\Gamma(M)=(1-4m^2/M^2)^q \cdot \Gamma_0$ where $q=1/2$ for
 $\sigma$-meson  (the $\pi$-mesons  are in $s$-wave; for $Z$-boson $q=1/2$ too)
 and $q=3/2$ for $\rho$- meson (the $\pi$-mesons are in $p$-wave).
 Taking $M=2m(1+x) << m_r$  we obtain
\be\label{U2m}
U_{2m}(r)=\frac{D_q g_1 g_2 \exp(-2mr)}{r(mr)^{q+1}}\\
D_q=\frac{2^{3/2-q}(2q)!!m_r m^2\Gamma_0}{(M_m^2(2m) -4m^2)^2} \sim
 \frac{m^2 \Gamma_0}{m_r (m_r-2m)^2}
\ee
In the estimate we assumed $M_m(2m)\approx m_r$.
This contribution has the same nature as the Uehling potential
produced by virtual ``decay'' of photon to electron-positron pair ($q=1/2$).
  
   Consider now the interval of the integration $2m<<M<<m_r$ which corresponds
to the distances $1/m_r <<r<<1/m$. The result of the integration in this area
 is
\be\label{Up}
U_{p}(r)=\frac{2g_1 g_2\Gamma(r^{-1})}{\pi m_r^3 r^3}
\ee
The potentials $U_{p} \propto 1/r^3$ ($1/m_r <<r<<1/m$)
 and $U_{2m}\propto \exp(-2mr)/r^{q+2}$ ($r>>1/m$)
cover the area $r >> 1/m_r$. The potential here is exponentially enhanced
 in comparison with the Yukawa potential- see  Fig.\ref{sigma}.
 In the region $ r \sim 1/m_r$ the
 resonance contribution
dominates. To estimate the correction to the Yukawa potential we present the
integral in Eq. (\ref{UBW}) in the following form:
\be\label{Ures}
U_{res}(r)=\frac{g_1 g_2 m_r\exp(-r m_r)}{ \pi r} I(r)\\
I(r)=\int^{\infty}_{2m}
\frac{\exp[r (m_r-M)] \Gamma_r 2M dM }{[(M^2-m_r^2)^2+m_r^2 \Gamma_r^2]} \\
\approx  \int^{\infty}_{0}
\frac{[1-r (M-m_r)] \Gamma_r 2M dM }{[(M^2-m_r^2)^2+m_r^2 \Gamma_r^2]}.
\ee 
This gives an estimate of the corrections to the
Yukawa potential
\be\label{Ures1}
U_{res}(r) \approx \frac{g_1 g_2}{r} \exp(-r m_r)
 (1- \Gamma_r r/\pi-\Gamma_r /(\pi m_r)).
\ee 
 Thus, to use Yukawa potential we must assume $\Gamma_r << \pi m_r$ and 
 $\Gamma_r r <<\pi$. Then we should add $U_{p}$ or $U_{2m}$
 in the area $m_r r >1$.

    The behaviour of the potential at small distances
depends on the convergence of the integral in the area $m<<M<1/r$.
 For mesons of finite size
the integral is convergent at $r=0$ and $U_{inf}(r)$ should be neglected.
The meson exchange theory of strong interaction is not applicable
at small distance anyway.

\begin{figure}
 \includegraphics[width=7cm]{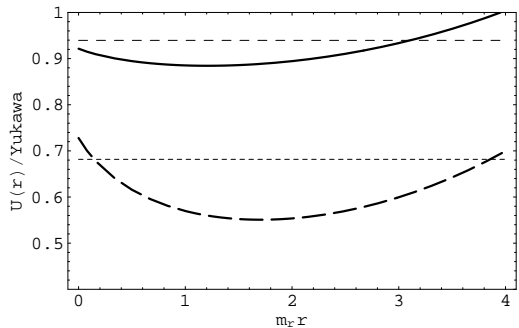}
 \caption{\label{sigm1}
The ratios of $\rho$-meson  potential to Yukawa potential
(thick solid line) and  $\sigma$-meson  potential to Yukawa potential
(thick dashed line) as functions of $m_r r$. Thin dashed lines
show factors  $1-\Gamma_r /(\pi m_r)$.}
 \end{figure}

Now we can compare our analytical estimates with numerical calculations
presented at  Fig.\ref{sigma}. To calculate the integral
 in Eq.(\ref{UBW}) we take\\
 $\Gamma(M)=[(1-4m^2/M^2)/(1-4m^2/m_r^2)]^q \cdot \Gamma_r$;\\
$\Gamma_r/m_r=0.19$,  $2m/m_r=0.36$ and  $q=3/2$ for $\rho$-meson;
$\Gamma_r/m_r=1$,  $2m/m_r=0.5$ and  $q=1/2$ for $\sigma$-meson.
The condition $\Gamma_r<<m_r$ is not satisfied
for  $\sigma$-meson. Nevertheless, the negative
corrections $1- \Gamma_r r/\pi-\Gamma_r /(\pi m_r)$ given in Eq.(\ref{Ures1})
 correctly explain behavior of 
both potentials at  $m_r r <1$
 (including the negative slope for $m_r r <0.5$, see Fig.\ref{sigm1} ).
 For $m_r r <3$ the  $\rho$-meson  potential
in  Eq.(\ref{UBW}) is about 10\% smaller than the Yukawa potenatial,
the  $\sigma$-meson potential  is about 40\% smaller.
 At $m_r r>5$ the positive corrections Eqs.(\ref{U2m}-\ref{Up})
give dominating contribution and change the long-range
behavior of the potentials\footnote{Note, however, that the potentials
are very small in this area.}.

  Finally, let us consider   potential
$U(r)$ in the short-range approximation,
\be\label{Udelta}
 U(r)\approx B \frac{4\pi g_1 g_2}{m_r^2}\delta(r),\\
B=\frac{m_r^2}{4\pi g_1 g_2}\int U(r)d^3r=\\
\frac{2}{\pi}\int^{\infty}_{2m/m_r}
\frac{\gamma(x) dx }{x [(x^2-1)^2+\gamma^2(x)]},
\ee
where $x=M/m_r$ and $\gamma(x)=\Gamma(M)/m_r$.
 This short-range approximation may also be useful in the calculation of the
nuclear mean field.
 The numerical integration gives $B=0.93$ for
$\rho$-meson and  $B=0.62$ for $\sigma$-meson.

 It may also be useful to have some analytical estimates for $B$.
In the logarithmic approximation
\be\label{B}
B \approx[1+
\frac{\Gamma_0}{\pi m_r}(\ln{\frac{m^2_r}{4m^2}}-const)].
\ee
It is interesting that the logarithmic correction diverges at $m=0$.
 This happens because
 of the long-range character of the potential
in the area  $1/m_r <<r<<1/m$. Indeed, the integration of
 $U_p(r) \propto 1/r^3$ from Eq.(\ref{Up}) immediately gives
 this logarithmic correction to $B$ . However,
for real mesons $\ln[(m_r/(2m)] \sim 1$ and the negative
 correction factor $1-\Gamma_r /(\pi m_r)$ to the resonance
contribution ( see  Eq.(\ref{Ures1}))
is more important. In fact,  numerical results for $B$ are surprisingly
close to this correction. 
Indeed,  $1-\Gamma_r /(\pi m_r)=0.94$  and $B=0.93$ for  $\rho$-meson; 
$1-\Gamma_r /(\pi m_r)=0.68$ and $B=0.62$ for  $\sigma$-meson.
 This gives a very
simple rough estimate for the effect of large width for the mesons:
multiply the Yukawa potential by the factor  $1-\Gamma_r /(\pi m_r)$.
The results of this prescription in comparison with real potentials
are shown at Fig.\ref{sigm1}.

\section{Long-range parity violating potential}
The $Z$-boson exchange creates a parity violating potential.
A possibility of a longer range of such potential could have
interesting applications, for example, in atomic tests of
the Standard Model (see e.g. review of atomic parity violation in
 \cite{ginges} and a different mechanism for a long-range parity
 violating potential in \cite{F}). 
  Note that the case of $Z$-boson differs from the meson case.
The imaginary part of the polarization operator for point-like
particles like photon and $Z$-boson rapidly increases with $M$. For example,
the electron contribution to the photon vacuum polarization is
 \cite{berestetskii}
\be\label{impol}
{\rm Im}P(M)=-\frac{\alpha}{3}(1-\frac{4m^2}{M^2})^{1/2}
 (2m^2+M^2)
\ee
For the $Z$-boson we need to sum over all decay channels, however, it does
not change the conclusions.
First consequence of ${\rm Im}P(M) \propto M^2$
 is that the high-energy contribution to the integral
(\ref{UBW}) gives famous running coupling constants (correction
$U_{inf}(r) \propto \ln{mr}/r$).
The second consequence is the strong dependence of the running width
  $\Gamma(r^{-1})$ on $r^{-1}$ in Eq. (\ref{Up}) since in this case
 $\Gamma (r^{-1})\sim \alpha (m^2 +r^{-2})/m_r$. Therefore,
the dominating contribution in the area $1/m_r <<r<<1/m$ is
 actually $U_{p}(r) \propto 1/M_Z^4r^5$.
As a result, the correction to 
 the short-range approximation, $U(r)=\frac{4\pi g_1 g_2}{m_r^2} \delta(r)$,
 remains finite
for $m=0$ (for example, for zero neutrino mass). However, 
the $Z-\gamma$ mixing via electron loop  gives a long-range
contribution (the second term below) to the parity violating electron-nucleus
 interaction 
\be\label{PV}
W=\frac{G}{2\sqrt{2}}\gamma_5 [-Q_w \rho(r) + 
Z \frac{2 \alpha (1-4 sin^2 \theta_w) m_e^2}{3 \pi^2 r} I(r)]\\
I(r)=
\int^{\infty}_{1} \exp(-2 x m_e r) \sqrt{x^2-1} (1 + \frac{1}{2 x^2})dx \,\,
\ee
Here the first term is the usual parity violating electron-nucleus interaction
(see e.g. \cite{ginges}); $G$ is the Fermi constant, $\gamma_5$ is the Dirac
matrix, $Q_w \approx -N +Z (1-4 sin^2 \theta_w) $ is the weak charge,
$Z$ and $N$ are numbers of protons and neutrons,
 $\theta_w$ is the Weinberg angle, $\rho(r)$ is the nuclear density
normalised to 1 ($\rho(r) \approx \delta(r)$), $m_e$ is the electron mass.
To obtain this potential we used the imaginary part of the polarization
operator from Eq.(\ref{impol}) and the Standard model values of
the interaction constants 
\footnote{Note, that this result takes into account that there is no 
$Z-\gamma$ mixing for zero momentum transfer, the polarization operator
$P(0)=0$ - see e.g.  \cite{berestetskii}.}. The integral has the following
asymptotics: for $m_e r \ll 1$
\be\label{PVsmall} 
I(r)=\frac{1}{(2 m_e r)^2}.
\ee
This gives $W \propto 1/r^3$. For $m_e r \gg 1$
\be\label{PVlarge} 
I(r)=\frac{\sqrt{\pi}\exp(-2  m_e r)}{4 (m_e r)^{3/2}}.
\ee
This gives $W \propto \exp(-2  m_e r)/r^{5/2}$.
 Thus, we have the parity violating potential with the range $1/(2 m_e)$.
A similar effect appears for the nuclear-spin-dependent part of
the parity violating interaction.  

{\bf Acknowledgments.\,\,}
This work was partially supported by the US-DOE grants DE-FG02-88ER40388
and DE-FG03-97ER4014, Australian Research Council and Gordon Godfrey fund.


\end{document}